\documentclass[aps,prb,twocolumn,superscriptaddress]{revtex4-1}

\usepackage{graphicx}
\usepackage{amssymb}
\usepackage{amsmath}
\usepackage{color}

\begin{document}

\title{Robustness of superconducting properties to transition metal substitution and impurity phases in Fe$_{1-x}$V$_{x}$Se}

\author{Franziska~K.~K.~Kirschner}
\email{franziska.kirschner@physics.ox.ac.uk}
\affiliation{Department of Physics, University of Oxford, Clarendon Laboratory, Parks Road, Oxford, OX1 3PU, United Kingdom}
\author{Daniel~N.~Woodruff}
\affiliation{Department of Chemistry, University of Oxford, Inorganic Chemistry Laboratory, South Parks Road, Oxford OX1 3QR, United Kingdom}
\author{Matthew~J.~Bristow}
\affiliation{Department of Physics, University of Oxford, Clarendon Laboratory, Parks Road, Oxford, OX1 3PU, United Kingdom}
\author{Franz Lang}
\affiliation{Department of Physics, University of Oxford, Clarendon Laboratory, Parks Road, Oxford, OX1 3PU, United Kingdom}
 \author{Peter~J.~Baker}
\affiliation{ISIS Facility, STFC Rutherford Appleton Laboratory, Chilton, Didcot, Oxfordshire OX11 0QX, United Kingdom}
\author{Simon~J.~Clarke}
  \affiliation{Department of Chemistry, University of Oxford, Inorganic Chemistry Laboratory, South Parks Road, Oxford OX1 3QR, United Kingdom}
\author{Stephen J.~Blundell}
\email{stephen.blundell@physics.ox.ac.uk}
\affiliation{Department of Physics, University of Oxford, Clarendon Laboratory, Parks Road, Oxford, OX1 3PU, United Kingdom}
\date{\today}

\begin{abstract}

We have performed transverse- and zero-field muon spin
rotation/relaxation experiments, as well as magnetometry measurements,
on samples of Fe$_{1-x}$V$_x$Se and their Li+NH$_3$ intercalates
Li$_{0.6}$(NH$_{2}$)$_{0.2}$(NH$_{3}$)$_{0.8}$Fe$_{1-x}$V$_{x}$Se. We
examine the low vanadium substitution regime: $x=0.005$, $0.01$, and
$0.02$. The intercalation reaction significantly increases the
critical temperature ($T_{\rm c}$) and the superfluid stiffness for
all $x$.  The non-intercalated samples all exhibit $T_{\rm c}\approx
8.5$~K while the intercalated samples all show an enhanced $T_{\rm
  c}>40$~K.  Vanadium substitution has a negligible effect on $T_{\rm
  c}$, but seems to suppress the superfluid stiffness for the
non-intercalated samples and weakly enhance it for the intercalated
materials.  The optimal substitution level for the intercalated
samples is found to be $x=0.01$, with $T_{\rm c} \approx 41$\,K and
$\lambda_{ab}(0) \approx 0.18$\,$\mu$m. The non-intercalated samples
can be modeled with either a single $d$-wave superconducting gap or
with an anisotropic gap function based on recent quasiparticle imaging
experiments, whereas the intercalates display multigap nodal behaviour
which can be fitted using $s+d$- or $d+d$-wave models.  Magnetism,
likely from iron impurities, appears after the intercalation reaction
and coexists and competes with the superconductivity. However, it
appears that the superconductivity is remarkably robust to the
impurity phase, providing a new avenue to stably improve the
superconducting properties of transition metal-substituted FeSe.

\end{abstract}

\maketitle

\section{Introduction}
The discovery of superconductivity in iron-based
systems\cite{Kamihara2006, Kamihara2008} has produced a range of new
high temperature superconductors. Among these compounds are those
based on FeSe, which in its undoped form \cite{Coldea2018} has a
critical temperature\cite{Hsu2008} $T_{\rm c} \approx 8$\,K.
Pressure,\cite{Medvedev2009} molecular
intercalation,\cite{Burrard-Lucas2013} and thin-film
fabrication\cite{Ge2015} can significantly enhance the
superconductivity in FeSe, with $T_{\rm c}$ reaching over 100\,K. The
studies on FeSe intercalates in particular have revealed a remarkable
robustness of the superconducting properties to structural
disorder.\cite{Foronda2015} Another common method for chemically
altering FeSe, and consequently enhancing superconductivity, is
through substitution on the chalcogenide site. Through tuning the
substitution fraction $x$ in FeSe$_{1-x}$S$_x$ and FeSe$_{1-x}$Te$_x$,
$T_{\rm c}$ increases from the FeSe value by 20\% and 75\%
respectively.\cite{Mizuguchi2009}

On the other hand, transition metal substitution of iron
(Fe$_{1-x}T_x$Se for transition metal $T$) has had more mixed
results. Superconductivity in Fe$_{1-x}$Cu$_x$Se is suppressed for
$x>1.5\%$, and $x>4\%$ drives the sample through a metal-insulator
transition.\cite{Williams2009} This is thought to occur owing to Cu
atoms disrupting the electronic structure, and eventually causing the
metal-insulator transition due to Anderson
localization.\cite{Chadov2010} Co and Ni substitution have been found
to either suppress\cite{Mizuguchi2009} $T_{\rm c}$ or destroy
superconductivity completely.\cite{Yadav2015} For $T=$ Mn, V, Cr, and
Ti, it has been found that $x$ can be tuned to
optimise\cite{Yadav2011, Yadav2013, Yadav2015} $T_{\rm c}$. An optimum
$T_{\rm c} \approx 11\,$K was found\cite{Yadav2015} for
Fe$_{0.98}$V$_{0.02}$Se.  It is thought that these highly
element-dependent results arise from both the ionic size and level of
impurity phases. Increased pressure on FeSe increases the fraction of
hexagonal impurity phase in the sample, which at first increases
$T_{\rm c}$, and then rapidly suppresses it.\cite{Medvedev2009} It has
been theorized that the amount of hexagonal phase could vary as a
function of chemical pressure which is related to the size of the
substituted transition metal ions,\cite{Yadav2015} although the
effect of chemical pressure from chalchogenide substitution in
FeSe$_{1-x}$(S,Te)$_x$ appears to be inconsistent with hydrostatic
pressure studies.\cite{Mizuguchi2009} A study on transition metal
substitution in FeSe$_{0.5}$Te$_{0.5}$ saw similar results to those
for FeSe and suggested that the differing magnetic properties between
the transition metal ions may induce different local impurity moments
and net carrier concentrations.\cite{Zhang2010} As a result of this
variation, it has also been suggested that the pairing symmetry of
transition metal substituted compounds may not be pure $s$- or
$d$-wave.

In this paper, we perform muon spin relaxation and rotation ($\mu$SR)
and magnetometry experiments on three samples of Fe$_{1-x}$V$_x$Se
(with $x=0.005$, $0.01$, and $0.02$) and their ammonia intercalates
(chemical formula
Li$_{0.6}$(NH$_{2}$)$_{0.2}$(NH$_{3}$)$_{0.8}$Fe$_{1-x}$V$_{x}$Se.;
labelled as $x+\rm{NH}_3$). Using transverse field (TF) $\mu$SR, we
extract the superconducting properties of all samples, and find that
the superfluid stiffness and critical temperature both increase
significantly after intercalation. We also observe superconductivity
with an anisotropic gap in both classes of samples, with the opening
of a second gap in the intercalates. Zero field (ZF) $\mu$SR and
magnetometry measurements reveal a strong magnetic signal in the
intercalates, which is absent in the non-intercalated samples, likely
arising from iron-based impurities. We find that the superconductivity
remains robust, despite the introduction of very small amounts of
elemental iron impurities produced during the intercalation reactions

\section{Experimental Details}

{\it Synthesis of Fe$_{1-x}$M$_x$Se:} iron powder (99.998\%, Alfa
Aesar), selenium powder (99.999\%, Alfa Aesar) and vanadium powder
($>$99.99\%) were ground together in the desired stoichiometry
in an agate mortar and pestle for 10 minutes before being sealed
inside a silica ampoule. This was heated to $700^\circ$C
($2^\circ$C/min) and kept there for 48~hours before being furnace
cooled to room temperature. The grey powder was then reground and
sealed inside a fresh silica ampoule. This was heated to
$700^\circ$C ($2^\circ$C/min) for 36~hours before being cooled to
400$^\circ$C and annealed for 10 days before being quenched to
$0^\circ$C. The isolated powders generally had small amounts ($<5$\%
by weight) of $\alpha$-FeSe impurity and were used as isolated for the
various intercalation reactions performed in this study.

{\it Synthesis of Li$_z$(NH$_3$)$_y$Fe$_{1-x}$V$_x$Se:} a similar
synthetic procedure was used to one we have reported
previously.\cite{Burrard-Lucas2013} A sample of Fe$_{1-x}$V$_x$Se
(500~mg) was placed inside a Schlenk tube along with lithium metal
($\approx$13~mg) and a Teflon-coated magnetic stirrer bar. The Schlenk
tube was evacuated and cooled to $-78^\circ$C using a
CO$_2$/isopropanol bath.  Ammonia ($\approx$15~ml) was condensed into
the flask to afford a dark blue solution. This was stirred for 4~hours
at $-78^\circ$C before the flask was allowed to warm to room
temperature naturally within the CO$_2$/isopropanol bath; all the
ammonia evaporated via a mercury bubbler. Once at room temperature the
flask was placed under dynamic vacuum for 2~minutes before the dark
grey material was then isolated inside an argon-filled glovebox.  The
intercalated samples exhibit sensitivity to air (see Appendix~A) and
precautions were taken to avoid exposure to air for these samples in
subsequent characterisation experiments.

{\it Diffraction Measurements:} X-ray powder diffraction (XRPD)
measurements were performed on instrument I11\cite{Thompson2009}
at the Diamond Light Source, with 0.826\AA\
X-rays and the position sensitive (MYTHEN) detector. Rietveld
refinements against powder diffraction data were conducted using the
TOPAS Academic software.\cite{topas}

{\it Magnetometry:} Field cooled (FC) and zero field cooled (ZFC)
magnetometry measurements were made using a Quantum Design MPMS SQUID
magnetometer which utilised measuring fields of 20--50\,Oe in order to
characterize the superconducting state and up to 7~T to probe the
normal state susceptibilities. Samples were sequestered from air in
gelatin capsules. Susceptibilities were corrected for the effect of
demagnetizing fields arising from the shape of the sample.

{\it Muon spin relaxation measurements:} $\mu$SR
experiments\cite{Blundell1999, Yaouanc2011} were performed using a
$^3$He cryostat mounted on the MuSR spectrometer at the ISIS pulsed
muon facility (Rutherford Appleton Laboratory, UK).\cite{King2013} TF
measurements, in which an external magnetic field is applied
transverse to the initial muon spin polarization, were made to
identify the superconducting ground state and its evolution with
$x$. ZF measurements were carried out on the $x=0.02+\rm{NH}_3$ sample
in order to test for magnetic phases in the sample. All of the $\mu$SR
data were analyzed using WiMDA.\cite{Pratt2000}

\section{Superconductivity} \label{SCsec}

\begin{figure*}[t]
	\includegraphics[width=.8\textwidth]{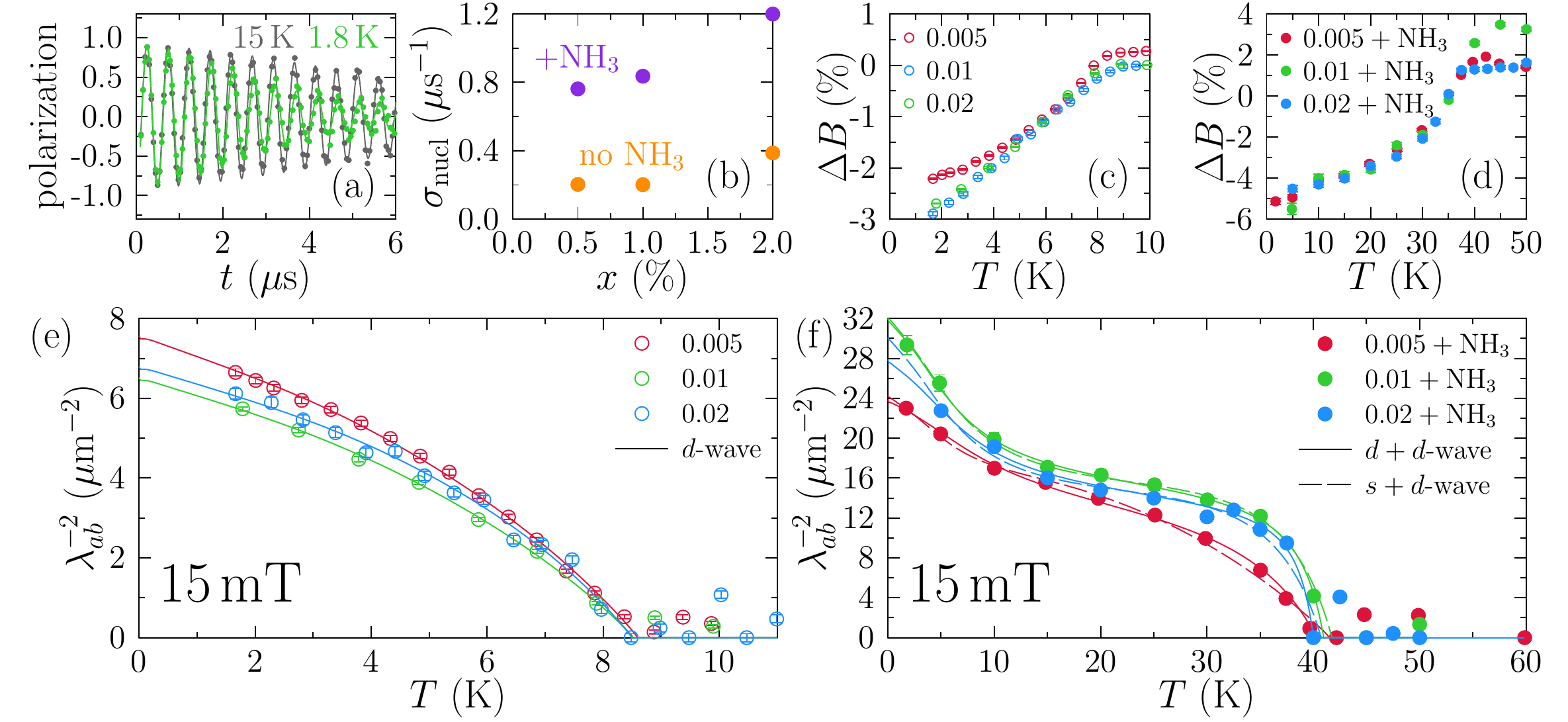}
	\caption{(a) Sample TF-$\mu$SR spectra above and below $T_{\rm
            c}$ for $x=0.01$. Fits as in Eq.~\ref{TFfit} are also
          plotted. (b) Dependence of the nuclear contribution to the
          superconducting relaxation (described in Eq.~\ref{TFfit}) on
          vanadium substitution for both intercalated ($+\rm{NH}_3$)
          and non-intercalated (no $+\rm{NH}_3$) samples. The
          temperature dependence of the field width of the
          superconducting vortex lattice is given in (c) and (d) for
          the non-intercalated and intercalated samples
          respectively. The temperature dependence of the inverse
          square penetration depth for non-intercalated and
          intercalated samples is shown in (e) and (f)
          respectively. The data in (e) have been fitted with a
          single-gap $d$-wave function, and the data in (f) have been
          fitted with two-gap $d+d$ and $s+d$ models.}
	\label{TFfig}
\end{figure*}

To determine the superconducting properties, all samples were measured
in a transverse field of $B_0 = 15\,$mT at temperatures $T$ above and
below $T_{\rm c}$. Sample spectra for $x=0.01$, as plotted in
Fig.~\ref{TFfig}a, show a clear increase in relaxation in the
superconducting state (compared to the normal state), arising from the
inhomogeneous magnetic field distribution of the vortex lattice. The
data were fitted with the two-component function
\begin{eqnarray} 
A(t) & = & A_{B} \cos \left( \gamma_{\mu} B_{0} t + \phi \right)
e^{-\lambda_{\rm TF} t} \nonumber \\  & & + A_{SC} \cos \left(
\gamma_{\mu} B_{\rm SC} t + \phi \right) e^{-\sigma^2 t^2/2},
\label{TFfit}
\end{eqnarray}
where $\gamma_{\mu} = 2\pi \times 135.5\,$MHzT$^{-1}$ is the
gyromagnetic ratio of the muon and $\phi$ is a phase related to the
detector geometry, with $\phi$ fitted for each of the eight detector
groups. The first term represents those muons which are not in the
superconducting volume and precess only in the external magnetic
field. These muons experience a small Lorentzian relaxation due to
magnetism in the sample (see below for further discussion), with
$\lambda_{\rm TF} \approx 0.1-0.2$\,$\mu$s$^{-1}$ showing little
variation between samples. The second term arises from muons in the
superconducting volume, which experience a Gaussian broadening
$\sigma(T) = \sqrt{\sigma_{\rm SC}^2(T) + \sigma_{\rm nucl}^2}$. This
broadening consists of a temperature-dependent component from the
vortex lattice, and a temperature-independent component from static
nuclear moments [plotted in Fig.~\ref{TFfig}(b)]. $\sigma_{\rm nucl}$ is
much higher for the intercalated samples, compared to the
non-intercalated samples which may reflect a contribution from static
{\it non-nuclear} (i.e.\ electronic) moments, although we note that
this contribution is temperature-independent.
	
The field shifts caused by the vortex lattice $\Delta B = B_{\rm SC} -
B_0$ in the non-intercalated and intercalated samples are shown in
Figs.~\ref{TFfig}(c) and (d), respectively. There is a clear negative
shift in the peak field as the samples transition into their
superconducting states; this is a characteristic feature of the vortex
lattice.\cite{Brandt1988}

In order to extract the penetration depth from $\sigma_{\rm SC}$, a conversion\cite{Brandt2003}
\begin{equation}
\sigma_{\rm SC} = 0.0609 \gamma_{\mu}\phi_0 \lambda_{\rm eff}^{-2}(T),
\end{equation}
was used; $\phi_0 = 2.069 \times 10^{-15}$\,Wb is the magnetic flux
quantum. All of the samples were anisotropic and polycrystalline and
so it can therefore be assumed that the effective penetration depth
$\lambda_{\rm eff}$ is dominated by the in-plane penetration depth
$\lambda_{ab}$, and so \cite{Fesenko1991} $\lambda_{\rm
  eff}=3^{1/4}\lambda_{ab}$. The temperature dependences of
$\lambda_{ab}^{-2}$ for the non-intercalated and intercalated
compounds are plotted in Figs.~\ref{TFfig}(e) and (f), respectively.

The data in Figs.~\ref{TFfig}(e) and (f) have been fitted with single-
and two-gap BCS models involving $s$- and $d$-wave gaps. The BCS model
of the normalized superfluid density of a superconductor is given
by:\cite{Chandrasekhar1993}
\begin{equation} \label{TFeq}
\tilde{n}_{\rm s}(T) =
\frac{\lambda_{ab}^{-2}(T)}{\lambda_{ab}^{-2}(0)} = 1 + \frac{1}{\pi}
\int^{2\pi}_{0} \int^{\infty}_{\Delta(\phi,T)} \frac{\partial
  f}{\partial E} \frac {E \,\rm{d}E \,\rm{d}\phi}{\sqrt{E^2 - \Delta^2
    (\phi,T)}},
\end{equation}
where $\Delta(\phi,T)$ is the superconducting gap function, and
$f=\left(1+\exp(E/k_{\rm B}T)\right)^{-1}$ is the Fermi function. The
gap function can be approximated as $\Delta(\phi,T) = \Delta(\phi)
\tanh \left(1.82 \left[ 1.018 \left(T_{\rm c}/T - 1 \right)
  \right]^{0.51} \right)$. The angular gap function
$\Delta(\phi)=\Delta_0$ for $s$-wave superconductors and $\Delta(\phi)
= \Delta_0 \cos (2\phi)$ for $d$-wave (nodal)
superconductors. Multi-gap systems can be represented by a sum of the
$\tilde{n}_{\rm s}(0)$ values for each individual gap, weighted by a
factor $w$ using
\begin{equation}
\tilde{n}_{\rm s}(T) = w\tilde{n}^{\rm (gap~1)}_{\rm
  s}(T) + (1-w)\tilde{n}^{\rm (gap~2)}_{\rm s}(T).
\label{eq:w}
\end{equation}

\begin{table*}[t]
	\centering
	
	\caption{Fitted parameters for the temperature dependence of
          $\lambda_{ab}^{-2}$ [plotted in Figs.~\ref{TFfig}(e) and
          (f)], using the fit in Eq.~\ref{TFeq}.}
	\label{TFtab}
\begin{tabular}{c c c c c c r c c c c}
	\hline\hline Sample & $a$ & $c$ & $c/a$ & Gap & $T_{\rm c}$ &
        $\Delta_1$ [symmetry] & $\Delta_2$ [symmetry] & $w$ &
        $\lambda_{ab}$\\ & (\AA) & (\AA) & & & (K) & (meV) & (meV) & &
        ($\mu$m)\\ \hline $0.005$ & 3.77076(3) & 5.52137(4) & 1.4643 &
        $d$ & 8.6(1) & 1.72(8) [d] & $-$ & $-$ & 0.36(1) \\ $0.01$ &
        3.77129(2) & 5.52105(3) & 1.4640 & $d$ & 8.5(3) & 1.71(15) [d]
        & $-$ & $-$ & 0.39(1) \\ $0.02$ & 3.77152(3) & 5.52164(6) &
        1.4640 & $d$ & 8.5(1) & 1.83(14) [d] & $-$ & $-$ & 0.38(1)
        \\ \hline $0.005 + \rm{NH}_3$ & 3.8315(1) & 16.3968(7) &
        4.2795 & $s+d$ & 41.1(8) & 0.62(1) [s] & 0.14(1) [d] & 0.61(9) &
        0.20(2) \\ & & & & $d+d$ & 40.2(9) & 1.1(1)[d] & 0.18(1) [d] &
        0.64(11) & 0.19(2) \\ $0.01 + \rm{NH}_3$ & 3.8336(1) &
        16.3429(4) & 4.2631 & $s+d$ & 41.7(6) & 1.23(5) [s] & 0.14(1)
        [d] & 0.45(5) & 0.18(1)\\ & & & & $d+d$ & 40.9(1) & 2.1(1)[d] &
        0.16(1) [d] & 0.53(4) & 0.18(1) \\ $0.02 + \rm{NH}_3$ &
        3.8295(1) & 16.4504(6) & 4.2957 & $s+d$ & 40.7(4) & 1.17(1)
        [s] & 0.14(1) [d] & 0.46(9) & 0.19(2)\\ & & & & $d+d$ & 40.0(1)
        & 2.3(1)[d] & 0.16(1) [d] & 0.55(9) & 0.20(2) \\ \hline\hline
	
\end{tabular}
\end{table*}

\begin{figure}[t]
	\includegraphics[width=.5\textwidth]{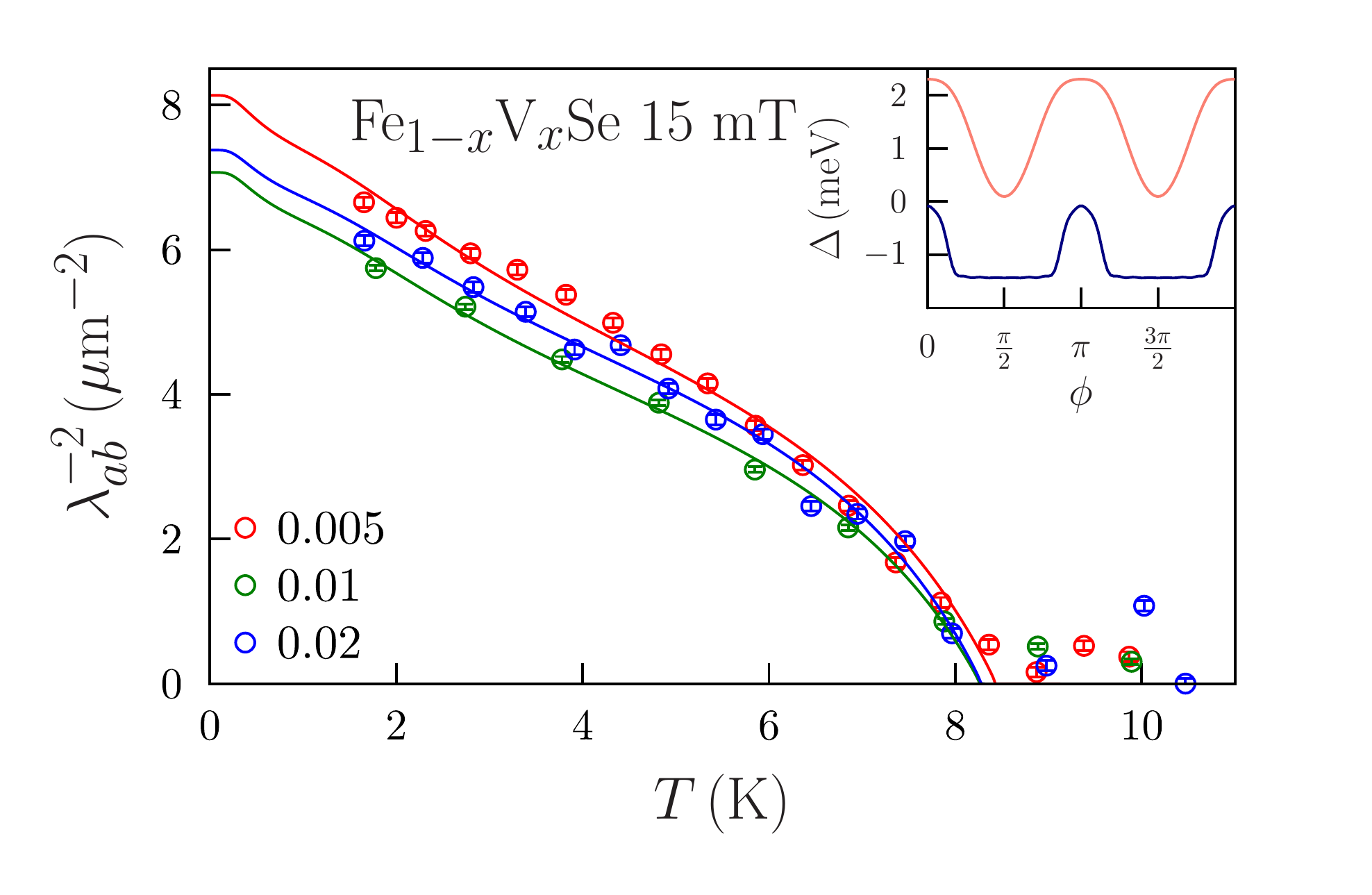}
	\caption{Fits of the data for Fe$_{1-x}$V$_x$Se to a gap
          function (see inset) based on the results of quasiparticle
          imaging experiments\cite{Sprau2017,Kreisel2017} on the
          surface of bulk FeSe, as described in the main text.}
	\label{fig:newfig}
\end{figure}

After trying combinations of $s$-wave and $d$-wave gap functions using
eqn~\ref{eq:w}. We find that the non-intercalated samples are best
described by a single-gap $d$-wave model (though we note the
sensitivity\cite{Sun2018} of the gap to disorder in FeSe) and the
extracted gap values (in the range 1--2~meV) are consistent with those
found for pure FeSe using other
techniques.\cite{Song2011,Kasahara2014,Sprau2017}  After
intercalation, an additional gap opens up: the intercalated samples
are described well by either $s$+$d$ or $d$+$d$ gaps (these two models
gave very similar $\chi^2$ values). The superconducting parameters
associated with the best fits are given in Table~\ref{TFtab}.
On the surface of pure FeSe, two gaps have been measured  using quasiparticle
interference imaging\cite{Sprau2017,Kreisel2017} and their angular
dependence mapped out (see Fig.~\ref{fig:newfig}).  We have used these
measured gap functions and eqn~\ref{eq:w} to fit the data on our
V-substituted FeSe samples and achieve a reasonable agreement with the
data (Fig.~\ref{fig:newfig}) and a very slightly lower estimate of the
penetration depth [$\lambda_{ab}(0)=$0.35(1), 0.37(1) and
  0.38(1)~$\mu$m for $x=$0.005, 0.01 and 0.02 respectively].
For the intercalated samples, a fit using the measured gap\cite{Zhang2016} for
monolayer FeSe was not successful (not shown).  This is likely due to
monolayer FeSe having a single Fermi surface
pocket\cite{Liu2012,Wang2012,Rebec2017} (as does
(Li,Fe)OHFeSe)\cite{Zhao2016} leading to a single gap with a more
conventional temperature dependence.\cite{BiswasLEM2018}  Although our
$\mu$SR data cannot help us pin down the pairing symmetry precisely,
it nevertheless provides strong evidence for two distinct gaps for the
ammonia-intercalated materials, probably resulting from the
additional pockets predicted for these compounds.\cite{Guterding2015} 

For the intercalated samples, we remark that the optimal substitution
level $x=0.01$ gives the largest value of $T_{\rm c}$ and the shortest
penetration depth (and therefore the largest superfluid stiffness,
which is proportional to $\lambda_{ab}^{-2}(0)$) though the variation
in both parameters as a function of $x$ is very slight. It has
previously been reported that optimum values of $x$ in transition
metal-doped FeSe exist,\cite{Yadav2015} above which $T_{\rm c}$
decreases, although for V-substitution, a previous study on
non-intercalated samples of Fe$_{1-x}$V$_x$Se found the optimal point
to be $x=0.02$.\cite{Yadav2015} For our non-intercalated samples, the
superconducting properties exhibit very little $x$-dependence (and
detailed $\mu$SR studies have been performed in pure FeSe
\cite{Khasanov2008,Biswas2018}).  We find, however, that transition
metal substitution significantly decreases the penetration depth,
compared to the undoped case ($\lambda_{ab}(0)\approx 0.41\,\mu$m and
$\approx 0.25\,\mu$m for the non-intercalated and intercalated FeSe
samples respectively).
	
\section{Magnetism} \label{magsec}

\begin{figure}[t]
	\includegraphics[width=.5\textwidth]{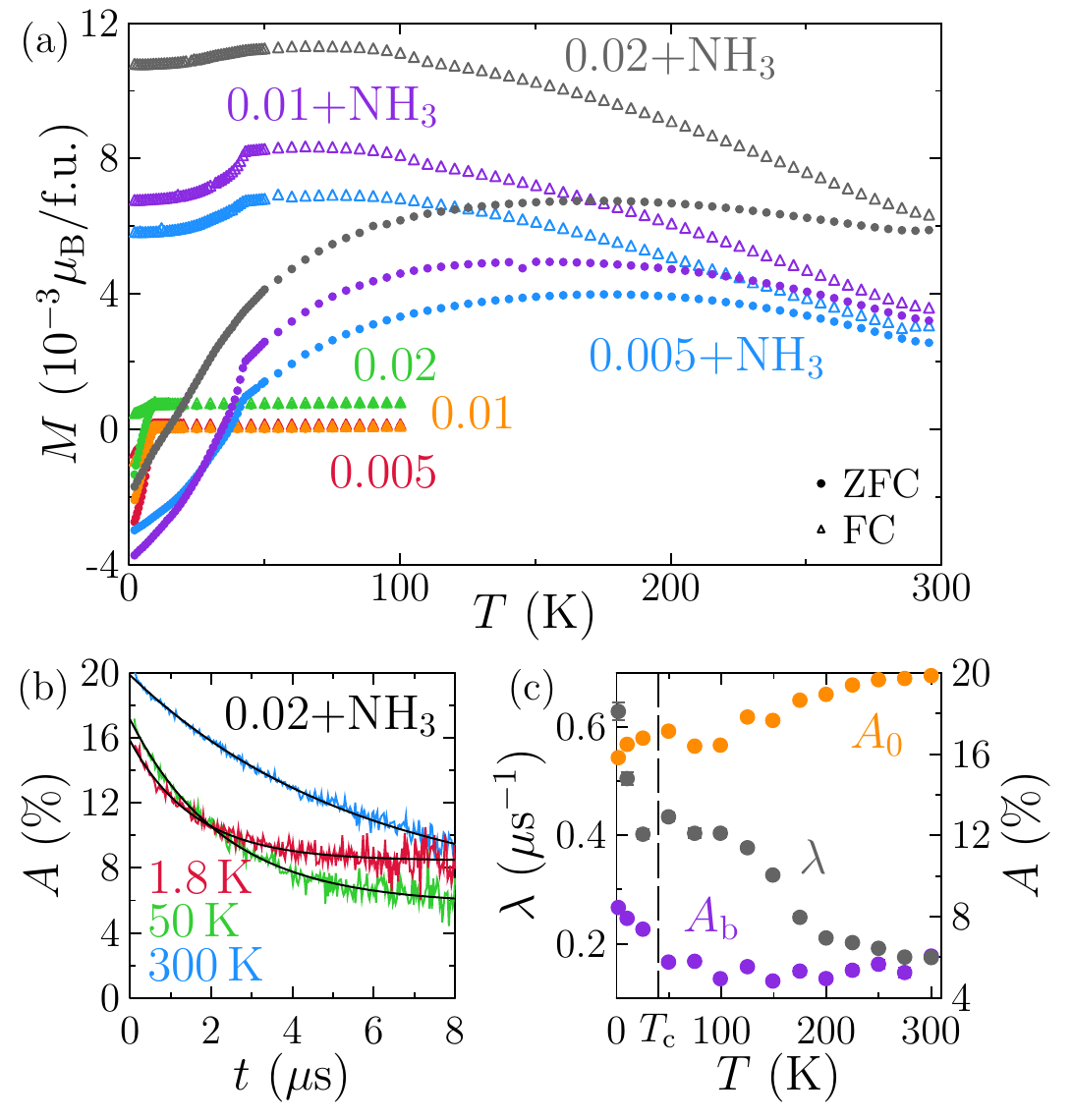}
	\caption{(a) Temperature dependence of the magnetisation for
          all samples, measured in Bohr magnetons per formula
          unit. (b) ZF-$\mu$SR asymmetry for $x=0.02+\rm{NH}_3$ at a
          range of temperatures. The black lines give fits as in
          Eq.~\ref{ZFfit} (c) Temperature dependence of relaxation
          $\lambda$, and initial and baseline asymmetries ($A_0$ and
          $A_{\rm b}$ respectively) of the ZF-$\mu$SR asymmetry for
          $x=0.02+\rm{NH}_3$.}
	\label{ZFfig}
\end{figure}

To examine the relaxation due to magnetism in the TF-$\mu$SR data, FC
and ZFC bulk magnetization measurements were carried out, and are
plotted in Fig.~\ref{ZFfig}(a). The non-intercalated samples were
found to undergo a superconducting transition at the expected
temperatures, with no strong magnetic signal above $T_{\rm
  c}$. However for the intercalated samples, we observe a clear
enhancement in the magnetization, which appears to arise from
elemental iron impurities. This enhancement overrides any
superconducting signal and, for $x=0.01+\rm{NH}_3$ and
$x=0.02+\rm{NH}_3$, the susceptibility is positive even well below
$T_{\rm c}$. From this we can conclude that the superconductivity
observed in the TF-$\mu$SR is likely strongly localized in a small
volume fraction

We performed ZF-$\mu$SR measurements on the $x=0.02+\rm{NH}_3$ sample,
to further investigate the bulk magnetic signal observed in the
magnetization data. Sample spectra well below, near, and above $T_{\rm
  c}$ are plotted in Fig.~\ref{ZFfig}(b). There appears to be no
Kubo-Toyabe relaxation, indicating that the magnetism is likely from
electronic moments rather than nuclear moments. The data were
well-modeled with a single-component Lorentzian relaxation:
\begin{equation} \label{ZFfit}
A(t) = \left(A_0 - A_{\rm b}\right) {\rm e}^{-\lambda t} + A_{\rm b},
\end{equation}
where $A_0$ and $A_{\rm b}$ are the initial and baseline asymmetries
respectively, and $\lambda$ is the relaxation rate. The fitted values
of $A_0$, $A_{\rm b}$ and $\lambda$ are plotted in
Fig.~\ref{ZFfig}(c).

As $T$ decreases, we find the initial asymmetry to decrease. This is
indicative of a fast-relaxing phase in the sample, which is outside
the resolution of the spectrometer. An increase in the baseline
asymmetry at low $T$ suggests a higher fraction of muons landing in
areas of the sample with no magnetic field (we note that $A_{\rm b}$
also contains a contribution from muons in the sample holder and
cryostat, but this contribution is expected to be
temperature-independent). One possible explanation of this behavior is
that the relaxation arises from magnetic puddles containing iron
impurities that freeze out at low temperatures to create areas of
static spin distributions with high resultant dipolar fields. This
change in the asymmetries and interpretation is consistent with the
increase in $\lambda$ in Fig.~\ref{ZFfig}(c). There are no
oscillations in the spectra in Fig.~\ref{ZFfig}(b), ruling out
long-range magnetic order.  We also find an increase in the magnitude
of the gradient of $A_0$, $A_{\rm b}$, and $\lambda$ below $\approx
T_{\rm c}$, which suggests the magnetism coexists and competes with
the superconductivity. Despite the strongly magnetic phase, it appears
that the superconductivity is robust to magnetism.
There is evidence that superconductivity in FeSe can be strongly affected by
the presence of disorder\cite{Bohmer2016} but our results show that
the presence of vanadium at low substitution levels produces
insufficient disorder to have a marked effect on $T_{\rm c}$.

\section{Conclusion}

We have performed TF- and ZF-$\mu$SR experiments, as well as
magnetization measurements, on three samples of Fe$_{1-x}$V$_x$Se and
their ammonia intercalates. In contrast with a previous
study,\cite{Yadav2015} we find that the optimal value of $x=0.01$
gives the highest critical temperature and superfluid stiffness,
although the dependence on $x$ is weak.  Another
study\cite{McQueen2009} has claimed that the amount of additional Fe
in interstitial sites between the FeSe layers reaches more than about
3\%, superconductivity can be destroyed.  There were no measurable
interstitial Fe ions in the Fe$_{1-x}$V$_x$Se phases according to the
X-ray diffraction measurements, consistent with the presence of
superconductivity (refined occupancy of $<1$\% with an uncertainty of
$\sim$1\%).  Intercalation increases these superconducting parameters
significantly, similar to that seen in pure FeSe and its
intercalate.\cite{Burrard-Lucas2013} The non-intercalated samples all
exhibit $T_{\rm c}\approx 8.5$~K while the intercalated samples all
show an enhanced $T_{\rm c}>40$~K.  Vanadium substitution has a
negligible effect on $T_{\rm c}$ but seems to suppress the superfluid
stiffness for the non-intercalated samples but enhance it for the
intercalated materials.  The non-intercalated samples can be modeled
with either a single $d$-wave superconducting gap or with an
anisotropic gap function based on recent quasiparticle imaging
experiments, whereas the intercalates display multigap nodal behaviour
which is best described using either $s+d$- or $d+d$-wave models.  In
the intercalation reactions with reducing sources of Li, the
thermodynamic products are Li$_2$Se and elemental Fe. In the reactions
with Li/NH$_3$ to obtain the products reported here, the intercalates
are metastable intermediates. As the susceptibility data show, some
elemental Fe is formed by partial decomposition at about the 5\% level
according to the magnetisation isotherms, but this does not destroy
the superconductivity in the intercalate phase.  The ZF-$\mu$SR
experiments suggest these impurities form localised magnetic regions,
which coexist and compete with the superconducting phase. In
Ref.~\onlinecite{Woodruff2016} we found that in some samples the
superconducting state co-existed with particles of expelled Fe, and
here we also find superconductivity is robust to the impurity
phase. This suggests that the line nodes in the intercalates are
likely symmetry-imposed and the impurity phase does not induce fully
gapped behavior.  An important drawback for the intercalated materials
however is that they are air-sensitive (see Appendix~A).  Our results
provide a novel route for creating intercalated FeSe compounds through
transition metal substitution on the Fe site.

\section{Acknowledgements}

We thank C.\ V.\ Topping for useful discussions. F.K.K.K.\ thanks
Lincoln College, Oxford, for a doctoral studentship. Part of this work
was performed at the Science and Technology Facilities Council (STFC)
ISIS Facility, Rutherford Appleton Laboratory and the Diamond Light
Source (beamtime allocation number EE13284 and EE18786).  We
acknowledge funding from EPSRC under grants EP/M020517/1 and
EP/N023803/1, Oxford's Centre for Applied Superconductivity (CfAS) and
the Leverhulme Trust under grant RPG-2014-221.

\appendix
\section{Air sensitivity}
\begin{figure}[htb]
	\includegraphics[width=.5\textwidth]{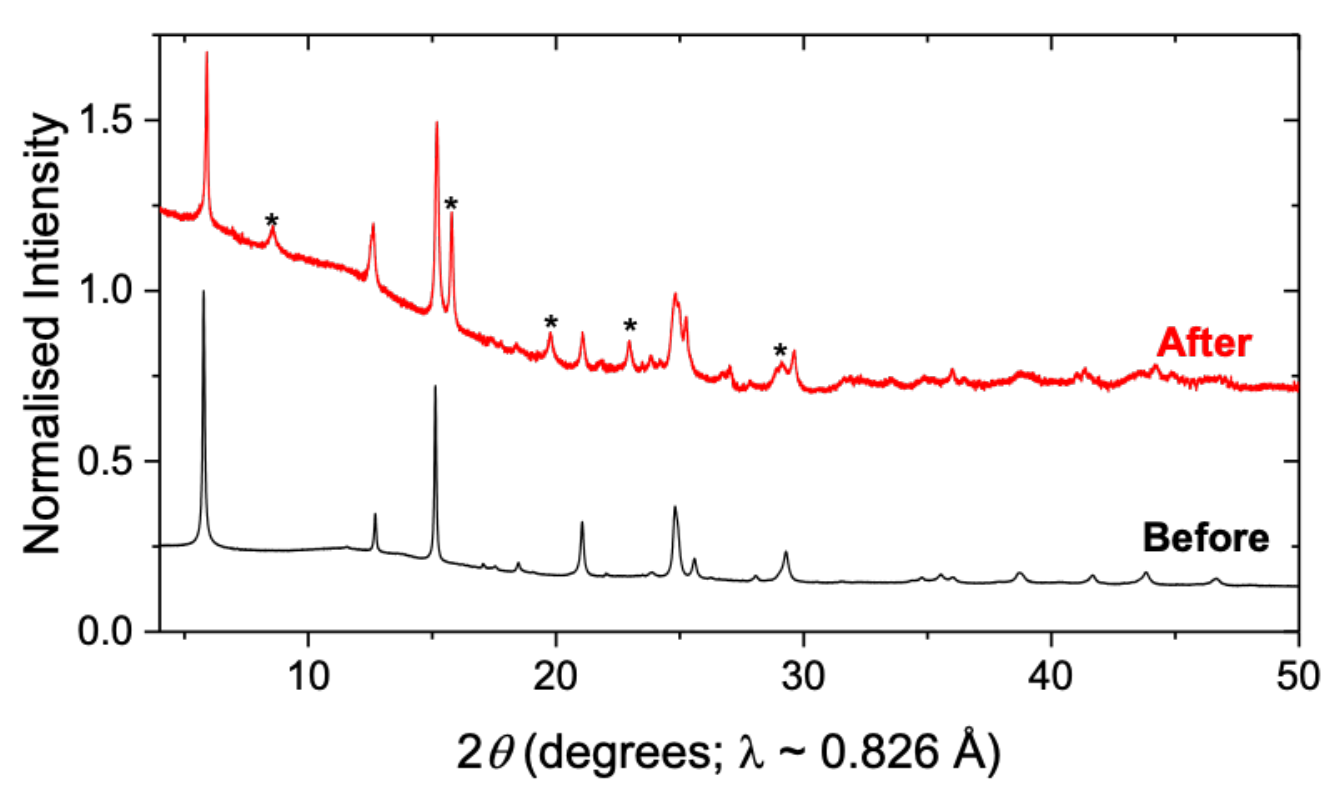}
	\caption{Powder diffraction data for the $x=0.005$ sample
          before and after exposure to air.  The asterisks indicate
          new impurity peaks.  Notice also the increase in the diffuse
          background.}
	\label{airexposure}
\end{figure}

The intercalated samples were stored under argon before each $\mu$SR
measurement and did not receive air exposure before being removed from
the muon spectrometer.  Following the $\mu$SR measurements, the
samples were kept wrapped in silver foil but were exposed to air for
several weeks.  They were then ground with glass (approximate 50:50
volume ratio) to avoid excessive absorption and preferred orientation
problems and packed into 0.5~mm diameter borosilicate capillaries. The
samples were then measured using powder X-ray diffraction at room
temperature using the Mythen position sensitive detector at the I11
beamline (Diamond, UK). A comparison of the powder diffraction
patterns for the $x=0.005$ sample taken before and after the $\mu$SR
measurement (i.e.\ before and after air exposure) is shown in
Fig.~\ref{airexposure}. Significant amounts of impurity phases
(indicated by asterisks in Fig.~\ref{airexposure}) have formed from
aerial decomposition of the product. However some of the intercalate
remains. This experiment was repeated for other compositions and in
some cases full sample decomposition had occurred, although we could
not ensure that each sample had received precisely the same amount of
exposure to air.  In any case, these results serve to demonstrate that
intercalated samples can suffer partial and potentially full
degradation when exposed to air for at least several days.  In contrast,
the non-intercalated samples are stable in air.
\bibliographystyle{apsrev4-1}
\bibliography{bib1}

\end{document}